\documentclass[12pt,preprint,referee]{aastex}

\RequirePackage{lineno}

\usepackage{txfonts}
\usepackage{gensymb}

\newcommand{\be} {\begin{equation}}

\newcommand{\ttt}[1]{\times10^{#1}}

\def\ae {AE Aqr}

\newcommand{\fermi}{{\em Fermi}}

\newcommand{\bc}{\begin{center}}
\newcommand{\ec}{\end{center}}
\def\ltsima{$\; \buildrel < \over \sim \;$}
\def\lsim{\lower.5ex\hbox{\ltsima}}
\def\loe{\lower.5ex\hbox{\ltsima}}
\def\gtsima{$\; \buildrel > \over \sim \;$}
\def\gsim{\lower.5ex\hbox{\gtsima}}
\def\goe{\lower.5ex\hbox{\gtsima}}

\def\ltsima{$\; \buildrel < \over \sim \;$}
\def\lsim{\lower.5ex\hbox{\ltsima}}
\def\loe{\lower.5ex\hbox{\ltsima}}
\def\gtsima{$\; \buildrel > \over \sim \;$}
\def\gsim{\lower.5ex\hbox{\gtsima}}
\def\goe{\lower.5ex\hbox{\gtsima}}

\def\ergscm2 {erg\,s$^{-1}$cm$^{-2}$}

\def\cm2 {cm$^{-2}$}

\shortauthors{Fermi-LAT collaboration}
\shorttitle{Fermi-LAT observations of \ae}

\begin{document}
\title{ Search for gamma-ray emission from AE Aquarii \\ with seven years of FERMI-LAT observations }

\author{Jian Li\altaffilmark{1}, Diego F. Torres\altaffilmark{1,2}, Nanda Rea\altaffilmark{1,3}, Emma de O\~na Wilhelmi\altaffilmark{1}, \\ Alessandro Papitto\altaffilmark{4}, Xian Hou\altaffilmark{5}, \& Christopher W. Mauche\altaffilmark{6} }
\altaffiltext{1}{Institute of Space Sciences (IEEC-CSIC), Campus UAB, Carrer de Magrans s/n, 08193 Barcelona, Spain}
\altaffiltext{2}{Instituci\'o Catalana de Recerca i Estudis Avan\c{c}ats (ICREA), E-08010 Barcelona, Spain}
\altaffiltext{3}{Anton Pannekoek Institute, University of Amsterdam, Postbus 94249, NL-1090-GE Amsterdam, The Netherlands}
\altaffiltext{4}{INAF-Osservatorio Astronomico di Roma, via di Frascati 33, I-00040 Monte Porzio Catone, Roma, Italy}
\altaffiltext{5}{Key Laboratory for the Structure and Evolution of Celestial Objects, Yunnan Observatories, Chinese Academy of Sciences, Kunming 650216, China.}
\altaffiltext{6}{Lawrence Livermore National Laboratory, L-473, 7000 East Avenue, Livermore, CA 94550, USA}

\begin{abstract}

AE Aquarii (AE Aqr) is a cataclysmic binary hosting one of the fastest rotating (P$_{\rm spin}$ = 33.08 s) white dwarfs known.
Based on seven years of \emph{Fermi} {Large Area Telescope (LAT)} Pass 8 data, we report on a deep search for gamma-ray emission from \ae\/.
Using X-ray observations from \emph{ASCA}, \emph{XMM-Newton}, \emph{Chandra}, \emph{Swift}, \emph{Suzaku}, and \emph{NuSTAR}, spanning 20 years, we substantially extend and improve the spin ephemeris of \ae\/.
Using this ephemeris, we searched for gamma-ray pulsations at the spin period of the white dwarf.
No gamma-ray {pulsations were} detected above 3 $\sigma$ significance.
Neither {phase-averaged} gamma-ray emission nor gamma-ray variability of \ae\ is detected by \emph{Fermi}-LAT.
We impose the most restrictive upper limit to the gamma-ray {flux} from \ae\ to date: $1.3\times 10^{-12}$ erg cm$^{-2}$ s$^{-1}$ in the 100 MeV$\textendash$300 GeV energy range, providing constraints on models.

\end{abstract}


\section{Introduction }
\label{intro}
Cataclysmic variables (CVs) are semi-detached binaries consisting of a white dwarf (WD) and a companion star, usually a red dwarf.
\ae\ is a bright {(V$\approx$11, Welsh et al. 1999)} CV hosting one of the fastest rotating WDs known (P$_{\rm spin}=33.08$ s, Patterson 1979) and a K 4-5 V secondary;
it is a non-eclipsing binary with {an orbital period} of 9.88 hr.
{Based on its trigonometrical parallax of $9.80\pm 2.84$ {mas} measured with {\em HIPPARCOS}, the distance to AE Aqr is estimated to be $102^{+42}_{-23}$ pc (Friedjung 1997; the large errors are due to the fact that AE Aqr is very faint for a {\em HIPPARCOS} measurement).}
\ae\ displays strong {broad-band} variability in the radio (Bookbinder \& Lamb, 1987; Bastian et al.\ 1988), optical (Beskrovnaya et al.\ 1996), ultraviolet (Eracleous et al.\ 1995, Mauche et al.\ 2012), X-rays (Oruru \& Meintjes 2012; Terada et al.\ 2008), and {possibly also at} TeV frequencies (Meintjes et al.\ 1994), although the latter has not been confirmed.
{The strength of the magnetic field of the white dwarf in \ae\ is uncertain, but based on the typical magnetic moments of intermediate polars (10$^{32}$ G cm$^{3}$) and polars (10$^{34}$ G cm$^{3}$), it is expected to lie in the {range B}$\sim$ 0.3--30 {$\times$ 10$^{6}$} G.
Specific estimates include B $\leq$ 2 {$\times$ 10$^{6}$} G (Meintjes 2002), B $\sim$ 1--5 {$\times$ 10$^{6}$} G (Cropper 1986; Stockman et al.\ 1992; Beskrovnaya et al.\ 1995), and B $\sim$ 50 {$\times$ 10$^{6}$} G (Ikhsanov 1998).}

Pulsations from \ae\ {at the spin period (P$_{\rm spin}$ = 33.08 s)} were first detected in the optical band (Patterson 1979), then confirmed in soft X-rays (Patterson et al.\ 1980), ultraviolet (Eracleous et al.\ 1994), and hard X-rays (Terada et al.\ 2008; Kitaguchi et al.\ 2014).
Radio pulsations were searched for with the Very Large Array, but only an upper limit {of 0.1 mJy was imposed} (Bastian et al. 1996).
{The spin-down power} of the white dwarf is $\dot{E}$=$-I \Omega\dot{\Omega}$ ($I \approx 2\times 10^{50}$ g cm$^2$ is the moment of inertia of {the} white dwarf, $\Omega $ and $\dot{\Omega} $ are respectively the spin frequency and its first derivative) {and is $6 \times 10^{33}$ erg s$^{-1}$ in the case of \ae\/}  (de Jager et al.\ 1994, {\ae\ is among the few CVs where $\dot{\Omega} $ is measured}).
This  exceeds the relatively low UV/X-ray luminosity ($L_{\rm UV} \sim L_{\rm X-ray} \sim 10^{31}$ erg s$^{-1}$) by about two orders of magnitude (see e.g., Oruru \& Meintjes 2012).

For the {high} magnetic field and the fast rotation period of the WD, \ae\ has been {characterized as a} ``white dwarf pulsar" (e.g. Bowden et al. 1992) and has been proposed to be a particle accelerator (Ikhsanov 1998; Ikhsanov \& Biermann 2006).
Thus, \ae\ could emit gamma-ray pulsations from a magnetospheric outer gap region, {similar to} gamma-ray pulsars (Abdo et al. 2013).
{Recently, the first radio pulsations in any white dwarf systems are detected in AR Scorpii (Marsh et al. 2016) and its broad-band spectrum is
characteristic of synchrotron radiation, which makes AR Scorpii another ``white dwarf pulsar".}

The lack of double-peaked emission lines combined with variations in line intensities associated with high velocities suggests the absence of a disc (e.g., {Welsh et al.\ 1998}).
Additionally, the mass-transfer rate from the secondary star{ ($\dot{M}$ $\sim$10$^{17}$ g s$^{-1}$, Wynn et al.\ 1997) is not high enough to overcome the magnetospheric pressure.}
For such mass-transfer rate, the accretion luminosity ($L_{\rm acc} = GM \dot M / R$, where $M$ and $R$ are the mass and radius of the WD) would be approximately three orders of magnitude larger than the observed X-ray luminosity.
Thus, \ae\ has been proposed to be in a magnetic propeller phase, ejecting most of the mass transferred from the secondary via the interaction with the magnetic field of the WD (Wynn et al.\ 1997).
This would be consistent with the fact that the rotation velocity at the magnetospheric radius for the quoted mass-transfer rate is much larger than the Keplerian velocity (see e.g., Oruru \& Meintjes 2012).

From this point of view, \ae\ could be a fast WD analog of the sub-luminous
state of transitional pulsars { in which the neutron star is surrounded by
an accretion disk but is sub-luminous in X-rays with respect to accreting
neutron stars (see e.g., Papitto \& Torres 2015 for a discussion)}, such as
IGR J18245-2452 (Papitto et al.\ 2013), J1023+0038 (Archibald et al.\ 2009;
Stappers et al. 2014; Patruno et al. 2014), and XSS J12270-4859 (e.g.,
Bogdanov et al.\ 2014, Papitto et al.\ 2015; de Martino et al. 2010, 2013).
For all these sources, gamma-ray emission has been found, and propeller
models have been developed to interpret it (e.g., Papitto, Torres, \& Li
2014, Ferrigno et al.\ 2014, Papitto \& Torres 2015, Campana et al. 2016).

Gamma-ray emission from \ae\ has been searched {for} at GeV and TeV energies.
Using observations spanning 4 years (1988--1991) with the Nooitgedacht Mk I Cherenkov telescope, at an average threshold energy $\sim 2.4$ TeV, Meintjes et al.\ (1992) {reported} a detection of pulsating TeV emission around the WD spin period.
Meintjes et al.\ (1994) reported simultaneous optical and TeV observations of \ae\ and confirmed the results of Meintjes et al.\ (1992).
TeV bursts with durations of minutes were also reported, but no pulsations were detected (Meintjes et al.\ 1994).
Later,  Bowden et al.\ (1992) and Chadwick et al.\ (1995) reported pulsating TeV emissions at the first harmonic of the WD spin period above 350 GeV with the Durham University {Very High Energy (VHE)} telescope.
{A 1-min long pulsating TeV burst} was {reported} from \ae\ by Bowden et al.\ (1992).
However, following these early observations, campaigns with the more sensitive Whipple Observatory (Lang et al.\ 1998) and the MAGIC telescopes (Aleksi{\'c} et al.\ 2014) have shown no evidence for any steady, pulsed, or episodic TeV emission of \ae\ {at} any epoch.
A negative result of a search for gamma-ray emission from \ae\ in the 0.1--1 GeV band was achieved with the Compton Gamma-ray Observatory {(Schlegel et al.\ 1995)}.
In this paper, we report on the search for gamma-ray emission and pulsations from \ae\ using a refined spin ephemeris of the white dwarf in \ae\/, using more than seven years {of the latest version of} \emph{Fermi}-LAT data\footnote{A \emph{Fermi}-LAT study of \ae\ with an earlier data release was presented in van Heerden \& Meintjes (2015). }.

\section{Observations}
\label{obs}

The \emph{Fermi}-LAT data included in this paper span seven years, from 2008 August 4 to 2015 August 7.
The analysis of the \emph{Fermi}-LAT data was performed using the \emph{Fermi} Science Tools,\footnote{\url{http://fermi.gsfc.nasa.gov/ssc/}} 10-00-05 release.
Events from the ``P8 Source'' event class {(evclass=128) and ``FRONT$+$BACK'' event type (evtype=3)} were selected\footnote{\url{http://fermi.gsfc.nasa.gov/ssc/data/analysis/documentation/Pass8$\_$usage.html}}.
The ``Pass 8 R2 V6'' instrument response functions (IRFs) were included in the analysis.
All photons within an energy range of 100 MeV--300 GeV and within a circular region of interest (ROI) of $10\degree$ radius centered on \ae\ were considered.
To reject contaminating {gamma rays} from the Earth's limb, we selected events with zenith angle $< 90\degree$.

The gamma-ray flux and spectral results presented in this work were calculated by performing a binned maximum likelihood fit using the Science Tool \emph{gtlike}.
The spectral-spatial model constructed to perform the likelihood analysis includes Galactic and isotropic diffuse emission components {(``gll\_iem\_v06.fits", Acero et al. 2016, and ``iso\_P8R2\_SOURCE\_V6\_v06.txt", respectively\footnote{\url{http://fermi.gsfc.nasa.gov/ssc/data/access/lat/BackgroundModels.html}}) }as well as known gamma-ray sources within $15\degree$ of  \ae\/, {based on a preliminary seven year source list}.
The spectral parameters of these sources were fixed at the source list values, except for sources within $3\degree $ of our target.
For these sources, all the spectral parameters were left free.
In the phase-related analysis, {photons within a specific phase interval are selected. To account for it,} the prefactor parameter {of the sources} were scaled to the width of the phase interval.
The test statistic (TS) was employed to evaluate the significance of the gamma-ray fluxes coming from the sources.
{The Test Statistic is defined as TS=$-2 \ln (L_{max, 0}/L_{max, 1})$, where $L_{max, 0}$ is the maximum likelihood value for a model without an additional source (the ``null hypothesis") and $L_{max, 1}$ is the maximum likelihood value for a model with the additional source at a specified location.
The larger the value of TS {the more likely that an additional source is needed}.
{TS $>$ 25 was the threshold for inclusion in the preliminary seven year source list.}
%
TS maps in this paper are produced with the \textit{Pointlike} analysis package (Kerr 2010).

For the {X-ray} timing analysis, we derived X-ray light curves using data from \emph{Swift}, \emph{Suzaku}, and \emph{NuSTAR}.
For the \emph{Swift}/XRT {observations} included in our analysis, we selected Photon Counting (PC) data with event grades 0--12 in {the} 0.3--10 keV {energy range}.
Source events were accumulated within a circular region {centered on the source} with a radius of 30 pixels (1 pixel = 2.36 arcsec).
{Background events were accumulated within a circular, source-free region with a radius of 60 pixels.}
For \emph{Suzaku} observations of \ae, we used datasets processed {with} the software of the \emph{Suzaku} data processing pipeline version 2.1.6.16.
We selected data from XIS 0--3 in the 0.3--10 keV energy range.
Reduction and analysis of the data were performed following the standard procedure.\footnote{\url{https://heasarc.gsfc.nasa.gov/docs/suzaku/analysis/abc/}}
{The source photons were accumulated from a circular region with a radius of
1 arcmin.
The background region was chosen in the same field of view with the same radius in a {source-free} region.}
For \emph{NuSTAR} observations of \ae\/, we selected data from both FPMA and FPMB in the 3--10 keV energy range.
{Source/background events were accumulated within a circular region with a radius of 30 arcsec centered on the source/source-free region.}
The data were processed and screened in the standard manner using the \emph{NuSTAR} pipeline software\footnote{\url{http://heasarc.gsfc.nasa.gov/docs/nustar/analysis/}}, NuSTARDAS version 1.4.1, with the \emph{NuSTAR} calibration database (CALDB) version 20141107.

X-ray data analysis was carried out using HEAsoft version 6.16.\footnote{\url{http://heasarc.nasa.gov/lheasoft/}}.
The times of arrival of the X-ray photons were {corrected} to the solar system barycenter using the Chandra-derived coordinates ({$\alpha$ =} 20:40:09.185, {$\delta$ =} $-$00:52:15.08; J2000, Kitaguchi et al.\ 2014) of \ae\/, which have sub-arcsecond uncertainties.
{All times are measured in terrestrial time (TT) and the DE 200 planetary ephemeris is used.}

\section{Search for steady gamma-ray emission}

Using seven years of \emph{Fermi}-LAT data, the \textit{gtlike} analysis of \ae\ yielded a TS value of 0 in the 100 MeV--300 GeV energy range: steady gamma-ray emission of \ae\ was not detected (Fig.\ \ref{ts}, left panel).
{There is a {TS excess beyond the assumed background model in the bottom left corner of the TS} map shown in Figure 1, but it is not significant (TS $<$ 25).}
We calculated a 99\% {confidence level (CL)} flux upper limit {on the steady {flux} from \ae} of $1.3\times 10^{-12}$ erg cm$^{-2}$ s$^{-1}$, according to Helene's method (Helene 1983), assuming a photon index of $2.0$ in the 100 MeV--300 GeV energy band.
{Systematic} effects have been considered by repeating the upper limit analysis using modified IRFs that bracket the effective area and changing the normalization of the Galactic diffuse model artificially by $\pm$6\%.

By selecting photons of spin phases 0.9--1.2 (see below for details) covering {periods} of relatively high gamma-ray counts (details of which are provided below and in {Section \ref{timing}}), we searched for gamma-ray emission from \ae\/.
No detection was made.
A 99\% CL flux upper limit of $4.4\times 10^{-12}$ erg cm$^{-2}$ s$^{-1}$ is evaluated according to Helene's method (Helene 1983) in the 100 MeV--300 GeV energy range, again assuming a photon index of $2.0$.

The 3-month binned long-term light curve of \ae\ is shown in Figure \ref{ts}, right panel.
All data points are upper limits and no flux variability is detected.
To search for gamma-ray flares as short as the TeV flare reported by Meintjes et al.\ (1994), we extracted the photons within $0.5\degree$ of \ae\ and produced an exposure-corrected photometry light curve in 1-minute bins.
No gamma-ray flare is detected above the {1.5} $\sigma$ level.

\begin{center}
\begin{figure*}
\centering
\includegraphics[scale=0.32]{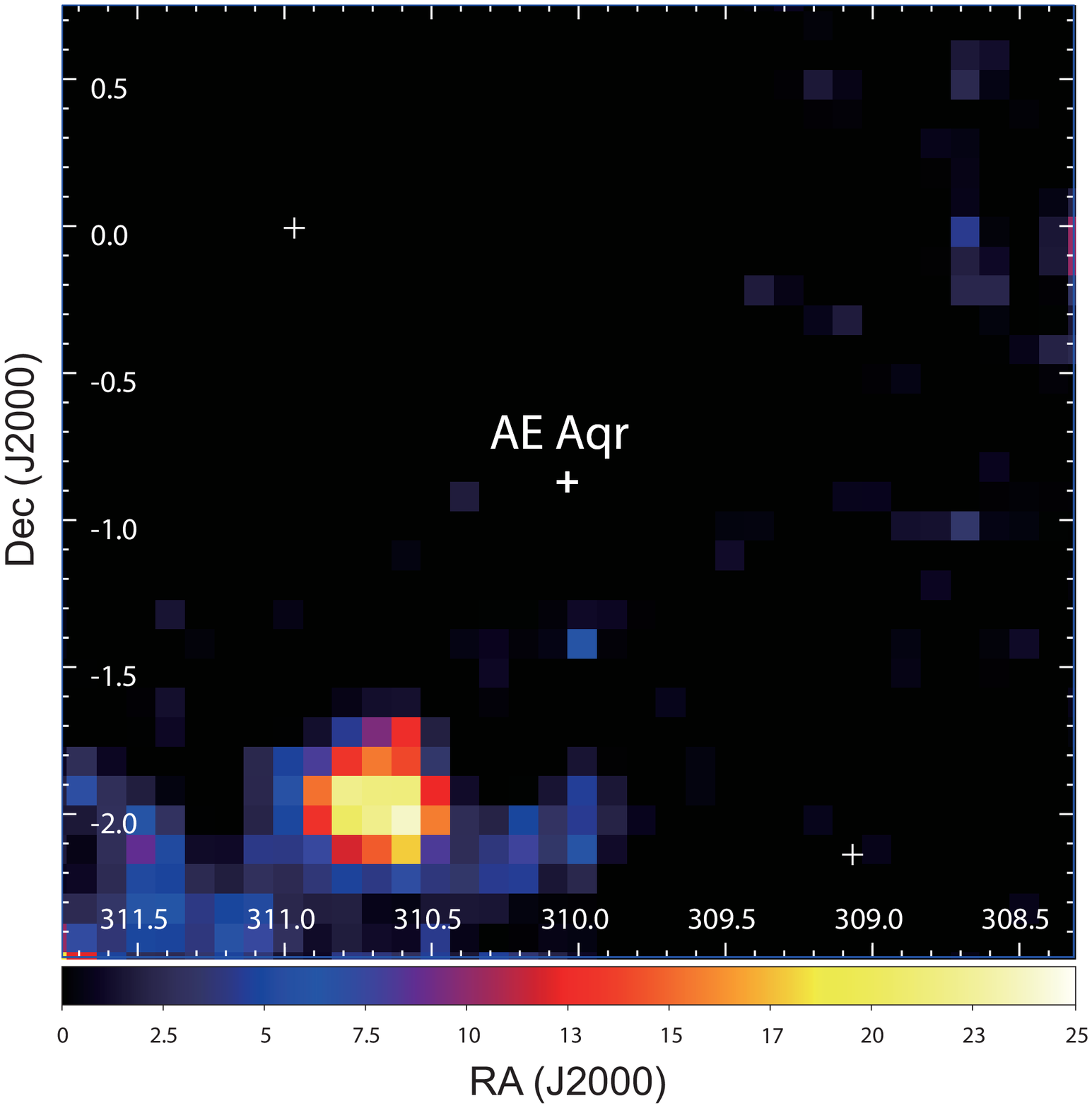}
\includegraphics[scale=0.42]{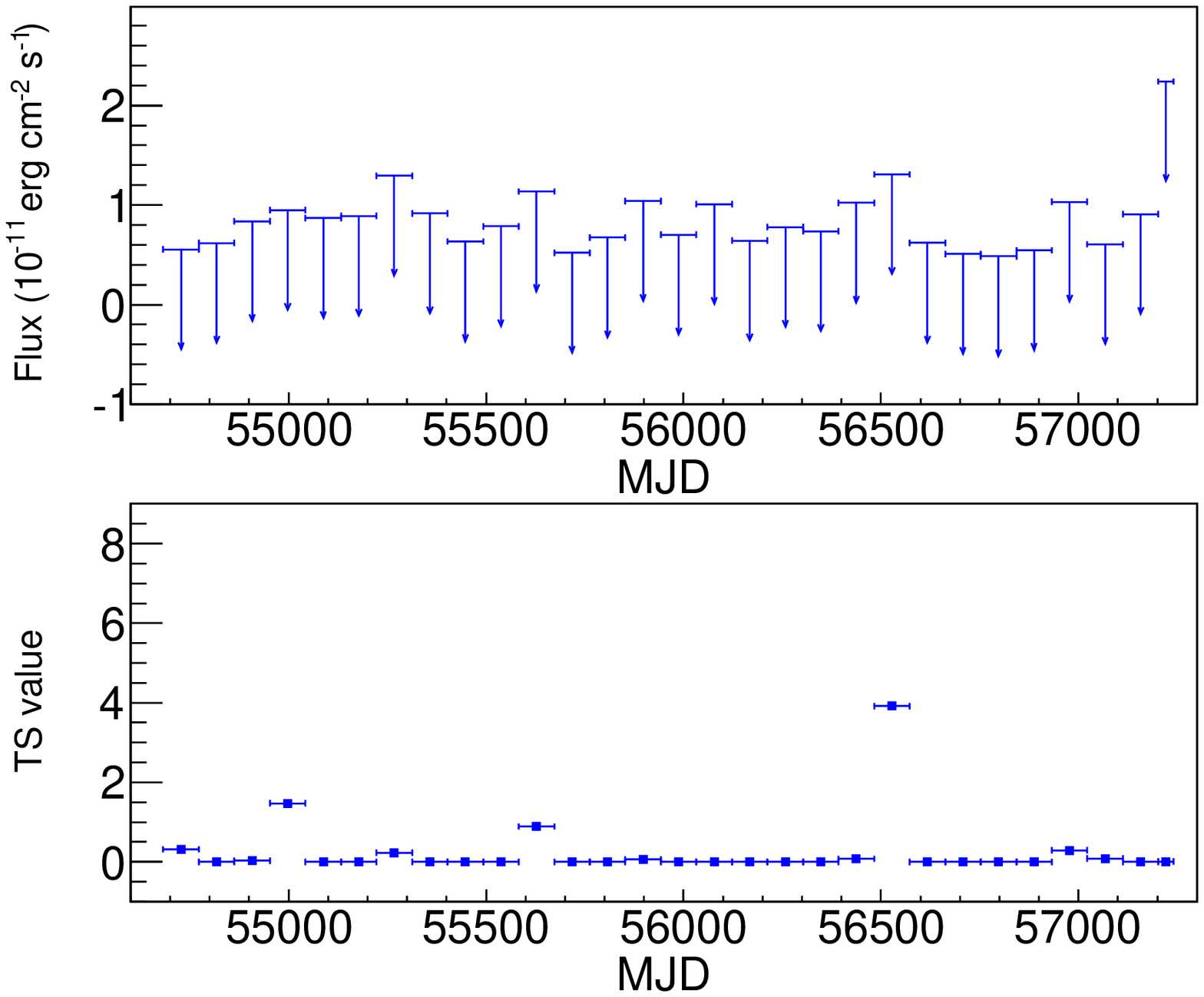}
\caption{Left: TS map of \ae\/ in the 100 MeV--300 GeV energy range.
The positions of \ae\/ and the sources in the {preliminary seven-year source list} are shown by the crosses;
%
%
{note} that the excess in the bottom left corner of the plot has a $TS < 25$, {and in any case is} far from the position of AE Aqr.
Right: 3-month binned long-term light curve of \ae\/. {Note that the last bin only spans 39 days.}}
\label{ts}
\end{figure*}
\end{center}

\section{Timing analysis {of} \ae\/}
\label{timing}


\begin{table*}{}
\centering
\scriptsize
\caption{X-ray observations and timing data}
\begin{tabular}{ccccccl}
\\
\hline
Date   & Observation duration &Spin phase offset  & Date of pulse maximum& $\chi^{2}$/d.o.f.  &  Satellite  \\
           &  (days)                          &                                   & (MJD)                                &                              &   \\

\hline\hline            
 1995 Oct.&  0.82    &0.037$\pm$0.011 &  50004.2037529$\pm$0.0000044	    &24.31/17 & \emph{ASCA}\\
 2001 Nov.&   0.15    &0.146$\pm$0.007 & 52221.0720142$\pm$0.0000029	    &28.22/17 & \emph{XMM-Newton}\\
 2005 Aug.&   0.82    &0.237$\pm$0.010 & 53612.7767058$\pm$0.0000039	    &18.53/17 & \emph{Chandra}\\
 2005 Aug.&  2.28     &0.274$\pm$0.060 &53612.7967672$\pm$0.0000229	     &1.05/3  & \emph{Swift}\\
 2005 Oct.&  2.08     &0.225$\pm$0.009 &53673.9503081$\pm$0.0000033	    &12.94/17 & \emph{Suzaku}\\
 2006 Oct.&   2.23    &0.263$\pm$0.009 &54033.2674939$\pm$0.0000036	    &25.43/17 & \emph{Suzaku}\\
 2009 Oct.&  3.46     &0.376$\pm$0.013 &55120.7649611$\pm$0.0000051	     &5.41/9  & \emph{Suzaku}\\
 2012 May &  32.88     &0.565$\pm$0.029 &56062.1941461$\pm$0.0000110	      &3.99/7  & \emph{Swift}\\
 2012 Sept.&  2.92     &0.537$\pm$0.013 &56174.8316376$\pm$0.0000051	    &40.60/17 & \emph{NuSTAR}\\
 2012 Sept.&   0.07    &0.521$\pm$0.069 &  56176.2607899$\pm$0.0000265	    &1.97/7  & \emph{Swift}\\
 2015 Jun.&   0.27    &0.630$\pm$0.041 &57176.1997047$\pm$0.0000159	    & 3.00/4  & \emph{Swift}\\
 2015 Dec.&   0.49   & 0.694$\pm$0.108 &57373.0316534$\pm$0.0000415	     &0.06/2  & \emph{Swift} \\
\hline\hline            
\label{xray}

\end{tabular}
\end{table*}

\label{results}

\begin{center}
\begin{figure}
\centering
\includegraphics[scale=0.8]{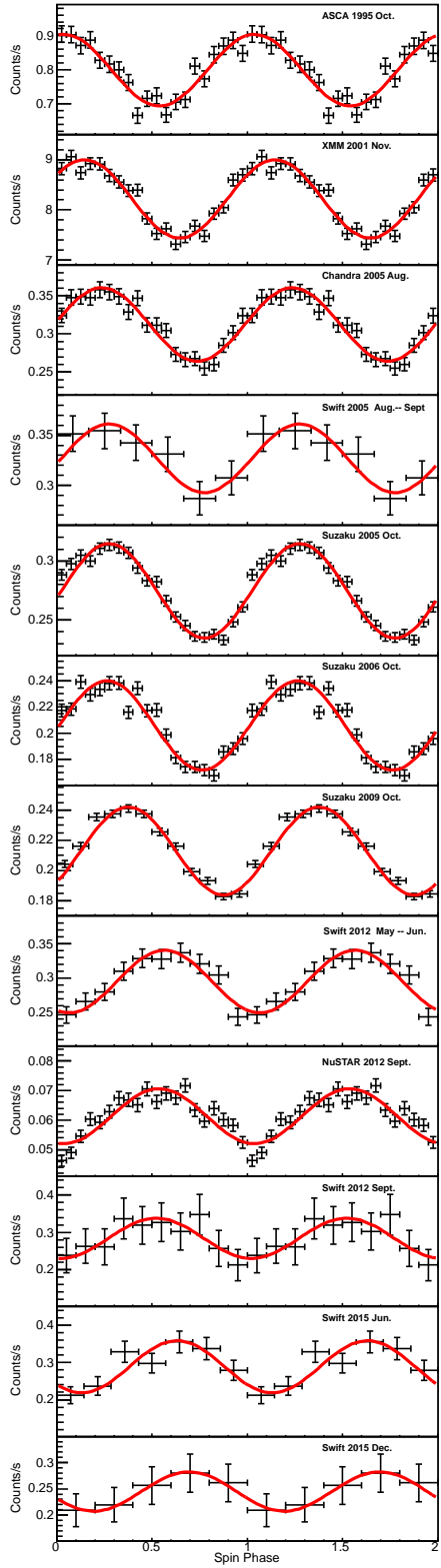}
\caption{Spin phase-folded X-ray light curves of \ae\ as obtained by the various satellites.
The \emph{ASCA}, \emph{XMM-Newton} and \emph{Chandra} light curves are taken from Mauche (2006).
{Two full rotations are shown for clarity.}
Best fitted {sinusoids} are shown by the solid red curves.}
\label{profile}
\end{figure}
\end{center}

\begin{center}
\begin{figure*}
\centering
\includegraphics[scale=0.8]{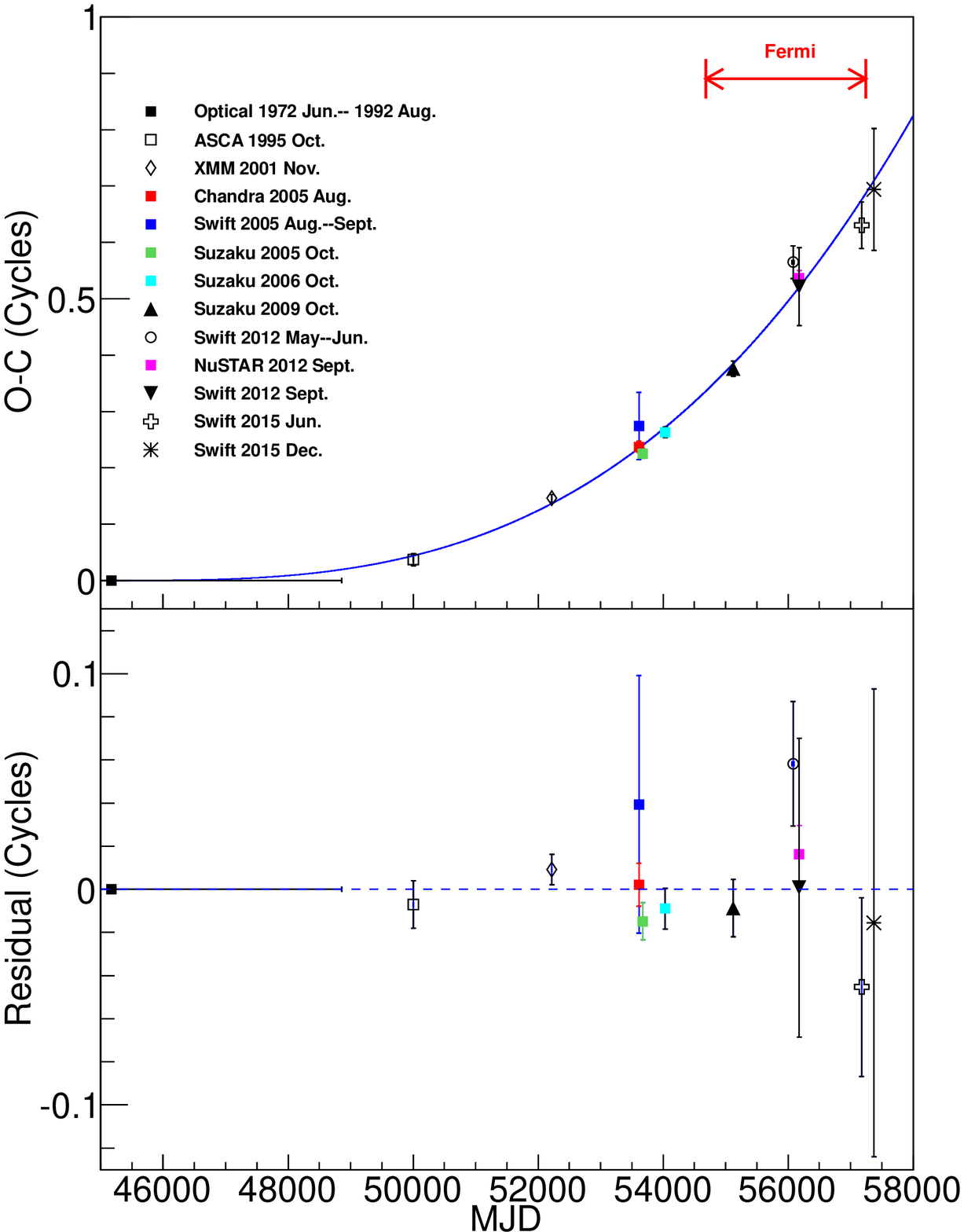}
\caption{Top: spin-phase offsets (Observed minus Calculated spin phases) of \ae\/ from different X-ray observations as a function of time.
The {function fit} to the data points is indicated by the blue curve.
The interval of \textit{Fermi} observations covered in this paper is shown by the red arrow.
Bottom: residuals of the fitting in above panel.}
\label{trend}
\end{figure*}
\end{center}

Rotational phases for each photon that passed the selection criteria could be calculated using TEMPO2 (Hobbs et al.\ 2006) with the \textit{Fermi} plug-in (Ray et al.\ 2011), and the significance of gamma-ray pulsations evaluated by the H-test (de Jager et al. 1989; de Jager \& B$\ddot{u}$sching 2010).
However, it is first necessary to have available a precise ephemeris of \ae\/ covering the span of the \emph{Fermi}-LAT data.
Using 14.5 years of optical data, de Jager et al.\ (1994) discovered that \ae\ is spinning down at a rate of $\dot{P} = 5.642(20)\times 10^{-14}$ {d d$^{-1}$}.
Relying on the fact that the optical and X-ray spin pulses are aligned in phase (Patterson et al.\ 1980), Mauche (2006) employed \emph{ASCA}, \emph{XMM-Newton}, and \emph{Chandra} X-ray observations spread over 10 yr to extend the baseline of observations of \ae\ to 27 years.
He found that the WD in \ae\ is spinning down slightly faster than given by the de Jager et al.\ ephemeris in a manner consistent with a second derivative of the period change, $\ddot{P} = 3.46(56)\times 10^{-19}$ d$^{-1}$.
Because the spin ephemeris is {\em descriptive} and not necessarily {\em predictive}, we added \emph{Swift}, \emph{Suzaku}, and \emph{NuSTAR} observations to extend and refine the spin ephemeris of \ae\/ over the span of the \emph{Fermi}-LAT observations (See Table \ref{xray}).
Due to the orbital motion of the WD around the binary center of mass, a photon arrival time delay of $\sim$ 2 seconds has been observed on \ae\/ {at optical (de Jager et al.\ 1994), ultraviolet (Eracleous et al.\ 1994), and X-ray (Mauche 2006) frequencies}.
As was done by Mauche (2006), before calculating the spin phases, we corrected the photon arrival times for the $2\cos(\phi_{\rm orb})$~second delays {for the orbital ephemeris of de Jager et al.\ (1994)}.
We produced spin phase-folded X-ray light curves of \ae\ for the various satellites assuming the de Jager et al.\ (1994) spin ephemeris; the light curves of \emph{ASCA}, \emph{XMM-Newton} and \emph{Chandra} are adopted from Mauche (2006).
Each light curve shows a similar sinusoidal profile and is fitted by a {sinusoid} function (Fig.\ \ref{profile}, red solid curves).
{The fitted $\chi^2$ values, degrees of freedom}, and spin phase offsets are listed in Table \ref{xray}.

It is evident from Figure \ref{profile} and Table \ref{xray} that the values of the spin-phase offsets derived from the various X-ray observations increase with time.
Figure \ref{trend} plots the spin-phase offsets versus time, demonstrating that the observed phases (``O'') diverge systematically from the calculated phases (``C'') assuming the de Jager et al.\ (1994) spin ephemeris.
{We fitted the data following the same method as Mauche (2006), yielding a new $\ddot{P} = 3.43(5)\times 10^{-19}$ d$^{-1}$ and a reduced $\chi^{2}$
of 1.14 (blue curve in Fig.\ \ref{trend}), which is consistent with, but reduces the error on, the $\ddot{P}$ term introduced by Mauche (2006).}
Thus, the spin ephemeris of \ae\/ extended for the \emph{Fermi}-LAT data is determined as:

$$
 \begin{array}{lll}
  \mbox{Epoch} & T_{0}  =  45171.500042\mbox{ (barycentered MJD)}  & \mbox{de Jager et al.\ (1994)}\\
  \mbox{Spin period } & P  = 0.00038283263840 \mbox{ d}& \mbox{de Jager et al.\ (1994)}\\
  \mbox{Spin period derivative } &\dot{P}=5.642\ttt{-14} \mbox{ d } \mbox{d}^{-1} & \mbox{de Jager et al.\ (1994)}\\
  \mbox{Spin period second derivative } & \ddot{P}' = 3.43 \ttt{-19}  \mbox{d}^{-1} & \mbox{this paper}\\
    \mbox{Orbital period} & P_{orbit} = 0.411655610 \mbox{ d } & \mbox{de Jager et al.\ (1994)}\\
    \mbox{Time of superior conjunction } & T'_{0} =45171.7784 \mbox{ (barycentered MJD)} & \mbox{de Jager et al.\ (1994)}\\
    \mbox{Projected semi-amplitude} & a_{WD} \sin i = 2.04 \mbox{ s} & \mbox{de Jager et al.\ (1994)}\\

\end{array}
 $$

Adopting this ephemeris, we searched for gamma-ray pulsations of \ae\ via an H-test procedure, for all \emph{Fermi}-LAT photons below 10 GeV, exploring different values {of} the minimum energy (from 100 MeV to 2.5 GeV in steps of 400 MeV) and radius ($0.3\degree$, $0.6\degree$, $0.9\degree$) {from \ae\/} of the \emph{Fermi}-LAT data.
The maximum {H-test} TS value we obtained is 18.8 in the 1.7--10 GeV energy range with {an} ROI radius of $0.6\degree $.
This is a 3.5 $\sigma$ result before trials correction, but only 2.6 $\sigma$ after considering the trials in minimum energy and {ROI radius}.
The folded pulse profile and TS value as a function of time is shown in Figure \ref{gamma_timing}, left and right panels.
{The H-test TS value shows an increasing trend as data {accumulate}.}

{We fitted a constant to the folded 1.7--10 GeV \emph{Fermi}-LAT spin gamma-ray light curve, {yielding an average flux of 6.92$\pm$0.83 counts and a reduced $\chi^{2}$ of 29.82/9}, demonstrating that the gamma-ray light curve is not consistent with being constant with spin phase.
A sinusoidal function plus a constant was also fitted to the gamma-ray light curve, yielding {a pulse
amplitude of 6.77$\pm$1.47 counts, a constant of 9.64 $\pm$ 1.12 counts, a spin phase offset of 0.01$\pm$0.03, and a reduced $\chi^{2}$ of 4.48/7.}
The F-test indicates that the fitting with a sinusoidal function plus a constant is favored over a constant fitting at 99.87\% confidence level {($\sim$ 3 $\sigma$).}
The folded 1.7--10 GeV Fermi-LAT spin gamma-ray light curve peaks around zero spin phase.
The optical, ultraviolet, and X-ray spin pulses of \ae\ are aligned in phase and also may be aligned with possible gamma-ray pulses if they {exist} just below the LAT sensitivity.}
However, no significant gamma-ray {pulsations} can be claimed at this time, because of the limited statistics.
The usual threshold for gamma-ray pulsar discovery is 5 $\sigma$ (Abdo et al. 2013).
\emph{Assuming \ae\ continues to follow the exact same pattern as it has shown in current \emph{Fermi}-LAT data},
the trial corrected significance of the gamma-ray pulsation from \ae\ will reach a 5 $\sigma$  level with an
additional $\sim$ 7.5 years of Fermi-LAT observations.
To track the evolution of the spin period and sustain the valid spin ephemeris, regular monitoring of \ae\ in the optical, ultraviolet, and/or X-rays, is required.

\begin{center}
\begin{figure*}
\centering
\includegraphics[scale=0.6]{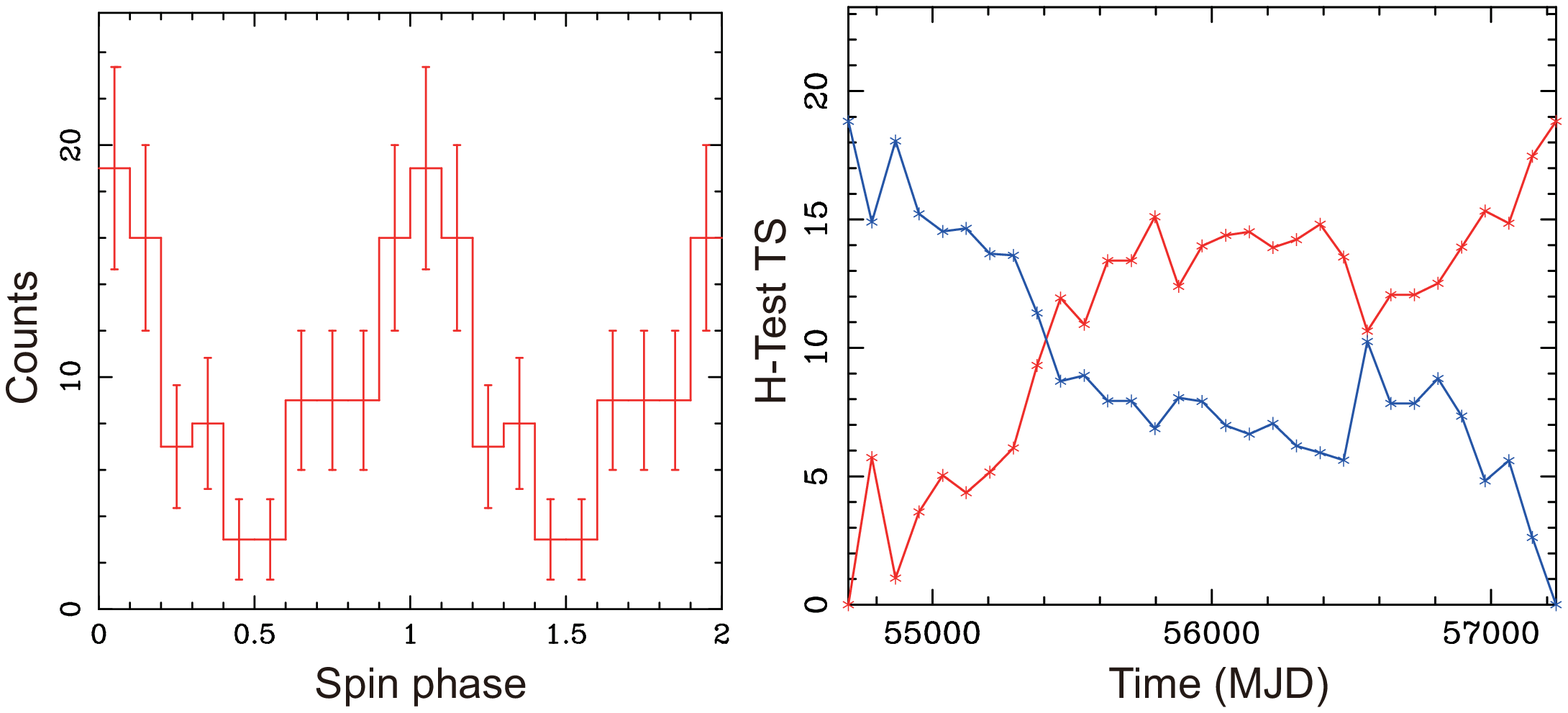}
\caption{Left, gamma-ray pulse profile of \ae\/ in the 1.7--10 GeV energy range, with a {ROI radius} of $0.6\degree$ folded on our timing ephemeris.
Two full rotations are shown for clarity.
Right, H-test TS value of \ae\/ as a function of time.
The red line represents the trend of TS value as data is accumulating from the start of the mission, while the blue line represents the same trend as data is accumulating from the end of the integration.}
\label{gamma_timing}
\end{figure*}
\end{center}

\section{Discussion}
\label{discussion}

We carried out the first detailed \emph{Fermi}-LAT data analysis of \ae\/, searching for gamma-ray pulsations and for steady gamma-ray emission.
Neither was detected.
We extended and refined the spin ephemeris of Mauche (2006) using data from 12 { different} observations by six X-ray satellites spanning 20 years.
Gamma-ray pulsations have been searched {for} using the extended ephemeris but no detection was made beyond a hint at the {2.6 $\sigma$ } significance level.
We calculated a 99\% CL flux upper limit {on the steady emission from} \ae\ of $1.3\times 10^{-12}$ erg cm$^{-2}$ s$^{-1}$ in the 100 MeV--300 GeV energy range.
This corresponds to a luminosity {upper limit} of $1.6\times 10^{30}$ erg s$^{-1}$ at 102 pc and {a gamma-ray efficiency of less than $2.7\times 10^{-4}$}.
No long-term flux variability is detected on a 3-month basis {and no flare activity was detected on {time} scales of 1 min}.

Meintjes \& de Jager (2000) proposed a propeller model to explain the TeV emission reported by Meintjes et al.\ (1994) and Chadwick et al.\ (1995), assuming that the magnetic field of the WD of \ae\ is rotating through a clumpy ring near the circularization radius.
In this model, particles could be accelerated to relativistic energies by the huge potential differences in the clumpy ring, leading to a gamma-ray luminosity $\sim 10^{34}$ erg s$^{-1}$ during bursts, which is equal to the total spin down power.
The upper limits to the steady gamma-ray luminosity measured by MAGIC ($\sim 6.8 \times 10^{30}$ erg s$^{-1}$ above 200 GeV, Aleksi{\'c} et al.\ 2014) {and our present analysis in the 100 MeV--300 GeV energy range are both several orders of} magnitude lower than this model prediction.
It is not likely that the model proposed by Meintjes \& de Jager (2000) could account for the gamma-ray emission of \ae\/, unless the episodes of gamma-ray emission are much more sporadic than expected, an assumption that seems ad hoc and unattractive.
{We note that the propeller model studied for transitional pulsars such as J1023+0038 discussed above was developed to explain a gamma-ray luminosity of the order of 10$^{34}$ erg/s, or $\sim$22\% of the corresponding pulsar's spin-down power.
The upper limit found in gamma rays is quite constraining in comparison to the spin down power of AE Aqr (it is similar to the limit reported by MAGIC, but at gamma-ray energies three orders of magnitude smaller).
We conclude that the rotation of the WD may be enough to preclude accretion, but not enough either to generate turbulence in the disc-field interface region, or reconnection events of the field, so that particles are accelerated up to TeV energies at a significant rate.}

The particle acceleration and energy release of a fast rotating, magnetic WD also could be explained in terms of the canonical spin-powered pulsar model.
With a pulsar-like acceleration process, Ikhsanov (1998) and Ikhsanov \& Biermann (2006) proposed the ejector white dwarf model (EWD) to explain the possible gamma-ray emission from \ae\/.
Applying the EWD model, the high energy emission of \ae\/ is dominated by the radiative loss of TeV electrons accelerated in the magnetosphere.
Gamma-ray emission would arise from inverse Compton scattering and the luminosity would be in the relatively uncertain range of 3--$500\times 10^{27}$ erg s$^{-1}$.
The upper limit derived in this paper is still one order of magnitude higher, so the EWD model prediction does not conflict with our results, although it is currently untestable.

\acknowledgments

The \fermi-LAT Collaboration acknowledges support from a number of agencies and institutes for both development and the
operation of the LAT as well as scientific data analysis. These include NASA and DOE in the United States,
CEA/Irfu and IN2P3/CNRS in France, ASI and INFN in Italy, MEXT, KEK, and JAXA in Japan, and the K.~A.\ Wallenberg
Foundation, the Swedish Research Council and the National Space Board in Sweden. Additional support from INAF in Italy and CNES in
France for science analysis during the operations phase is also gratefully acknowledged.

We acknowledge the support from the grants AYA2015-71042-P, SGR 2014-1073 and the National Natural Science Foundation of
China via NSFC-11473027, NSFC-11503078, NSFC-11133002, NSFC-11103020, XTP project XDA 04060604
and the Strategic Priority Research Program ``The Emergence of Cosmological Structures" of the Chinese Academy of Sciences, Grant No. XDB09000000.
N.R. is further supported by an NWO Vidi Award.
A.P. acknowledges support via an EU Marie Sklodowska-Curie Individual Fellowship under contract No. 660657-TMSP-H2020-MSCA-IF-2014, as well as fruitful discussion with the international team on ``The disk-magnetosphere interaction around transitional millisecond pulsars" at ISSI (International Space Science Institute), Bern.
C.W.M.'s contribution to this work was performed under the auspices of the US Department of Energy by Lawrence Livermore National Laboratory under Contract DE-AC52-07NA27344.
We acknowledge the assistance from Dr. Zhongli Zhang with the \emph{Suzaku} data analysis.
This research has made use of data obtained through the High Energy Astrophysics Science Archive Research Center Online Service, provided by the NASA/Goddard Space Flight Center.

\end{document}